# Encoding Enhanced Complex CNN for Accurate and Highly Accelerated MRI

Zimeng Li, Sa Xiao, Cheng Wang, Haidong Li, Xiuchao Zhao, Caohui Duan, Qian Zhou, Qiuchen Rao, Yuan Fang, Junshuai Xie, Lei Shi, Fumin Guo, Chaohui Ye, and Xin Zhou

*Abstract*—Magnetic resonance imaging (MRI) using hyperpolarized noble gases provides a way to visualize the structure and function of human lung, but the long imaging time limits its broad research and clinical applications. Deep learning has demonstrated great potential for accelerating MRI by reconstructing images from undersampled data. However, most existing deep conventional neural networks (CNN) directly apply square convolution to k-space data without considering the inherent properties of k-space sampling, limiting k-space learning efficiency and image reconstruction quality. In this work, we propose an encoding enhanced ($EN^2$) complex CNN for highly undersampled pulmonary MRI reconstruction. $EN^2$ employs convolution along either the frequency or phase-encoding direction, resembling the mechanisms of k-space sampling, to maximize the utilization of the encoding correlation and integrity within a row or column of k-space. We also employ complex convolution to learn rich representations from the complex k-space data. In addition, we develop a feature-strengthened modularized unit to further boost the reconstruction performance. Experiments demonstrate that our approach can accurately reconstruct hyperpolarized $^{129}$Xe and $^1$H lung MRI from 6-fold undersampled k-space data and provide lung function measurements with minimal biases compared with fully-sampled images. These results demonstrate the effectiveness of the proposed algorithmic components and indicate that the proposed approach could be used for accelerated pulmonary MRI in research and clinical lung disease patient care.

## I. INTRODUCTION

MAGNETIC resonance imaging (MRI) is a non-invasive and ionizing radiation-free imaging method that can provide exquisite structural and functional information [1]. Gas MRI can offer anatomical and functional information of the lung [2], [3], which provides unique opportunities for early diagnosis and treatment of lung disease [4-6]. However, long imaging time leads to image quality degradation because the longitudinal magnetization of hyperpolarized noble gases decays over time [7]. Accordingly, accelerating the acquisition speed is of great significance for clinical application of pulmonary gas MRI.

Several methods have been proposed to accelerate pulmonary gas MRI. Hardware-based solutions allow parallel imaging by sampling k-space data using multiple coils [8]. Pulse sequence enables fast image acquisition by optimizing the k-space sampling trajectory [9], [10]. Compressed sensing (CS) exploits the sparsity of signals in a specific transform domain to reconstruct high-quality images from undersampled k-space data [11-14].

Deep learning (DL) has recently become the focus of significant interest for MRI reconstruction [15]. Conventional reconstruction frameworks have been combined with deep learning models in various ways [16]. For example, Akçakaya et al. replaced the linear estimation in GRAPPA with convolutional neural networks (CNNs) [17], and Hammernik et al. employed variational networks to generalize the CS frameworks [18]. These methods require the integration of priors into the network design. Moreover, end-to-end networks can directly learn the mapping between the undersampled and fully sampled MRI data without the need for any priors, which exhibit superior performance in both reconstruction quality and speed [19-21]. A few studies applied complex-valued networks for MRI reconstruction considering the complex nature of MRI data [22,23]. Their results demonstrated that complex networks can more accurately reconstruct the phase and magnitude images compared with real-valued networks. Inspired by these important works, several studies introduced deep learning

This work was supported by National key Research and Development Project of China (2018YFA0704000), National Natural Science Foundation of China (82127802, 21921004, 82001915), Key Research Program of Frontier Sciences (ZDBS-Y-JSC004), Hubei Provincial Key Technology Foundation of China (2021ACA013), and China Postdoctoral Science Foundation (2020M672454, 2020T130125ZX). Xin Zhou acknowledges the support from the Tencent Foundation through the XPLORER PRIZE

Z. Li, and Y. Fang are with the State Key Laboratory of Magnetic Resonance and Atomic and Molecular Physics, National Center for Magnetic Resonance in Wuhan, Wuhan Institute of Physics and Mathematics, Innovation Academy for Precision Measurement Science and Technology, Chinese Academy of Sciences – Wuhan National Laboratory for Optoelectronics, Wuhan 430071, China, also with School of Physics, Huazhong University of Science and Technology, Wuhan 430074, China (e-mail: zm_li@hust.edu.cn; fuangyuann@gmail.com).

S. Xiao, C. Wang, H. Li, X. Zhao, C. Duan, Z. Qian, Q. Rao, J. Xie, L. Shi are with the State Key Laboratory of Magnetic Resonance and Atomic and Molecular Physics, National Center for Magnetic Resonance in Wuhan, Wuhan Institute of Physics and Mathematics, Innovation Academy for Precision Measurement Science and Technology, Chinese Academy of Sciences - Wuhan National Laboratory for Optoelectronics, Wuhan 430071, China, also with the University of Chinese Academy of Sciences, Beijing, 100049, China (e-mail: xiaosa@wipm.ac.cn; wangcheng@apm.ac.cn; haidong.li@wipm.ac.cn; xiuchao.zhao@apm.ac.cn; duancaohui@163.com; zqaggg1994@gmail.com; d201677758@hust.edu.cn; junshuaixie@apm.ac.cn; lshi@apm.ac.cn).

F. Guo is with Wuhan National Laboratory for Optoelectronics, Department of Biomedical Engineering, Huazhong University of Science and Technology, Wuhan 430074, China (e-mail: fguo@hust.edu.cn).

C. Ye and X. Zhou are with the State Key Laboratory of Magnetic Resonance and Atomic and Molecular Physics, National Center for Magnetic Resonance in Wuhan, Wuhan Institute of Physics and Mathematics, Innovation Academy for Precision Measurement Science and Technology, Chinese Academy of Sciences - Wuhan National Laboratory for Optoelectronics, Wuhan 430071, China, also with the University of Chinese Academy of Sciences, Beijing, 100049, China, and also with Key Laboratory of Biomedical Engineering of Hainan Province, School of Biomedical Engineering, Hainan University, Haikou 570228, China (e-mail: ye@wipm.ac.cn; xinzhou@wipm.ac.cn). (Corresponding author: Xin Zhou.) (Zimeng Li and Sa Xiao contributed equally to this work.)

strategies to pulmonary gas MRI reconstruction. Duan and coworkers first proposed a cascaded CNN model (CasNet) to accelerate pulmonary gas MRI [24] and later developed a deep cascade of residual dense network (DC-RDN) for diffusion-weighted gas MRI reconstruction [25].

Although promising, there are still some challenges in using deep learning for accelerating pulmonary gas MRI. For example, the existing networks (CasNet and DC-RDN) mainly focus on de-aliasing in the image domain without utilizing raw k-space data. Several studies on $^1$H MRI reconstruction have demonstrated that k-space-based learning [26] and dual-domain (k-space and image-domain) learning can effectively improve reconstruction performance [27]. Current networks apply the commonly used square convolution kernel both in k-space and image domain learning. These methods ignore the characteristics of k-space data generated through phase-encoding and frequency-encoding processes, whereby each data point contains information about the whole MR image [28]. This is potentially problematic because there is little spatial correlation between the data within a square region of k-space. Therefore, conventional square convolution kernel may not be optimal for k-space learning. In addition, CasNet and DC-RDN only consider the magnitude information of pulmonary gas MRI data in the reconstruction process, which have limited capability to learn the representations of the complex-valued MRI data. Furthermore, CasNet and DC-RDN mainly perform image reconstruction at an acceleration factor (AF) of 4 and the reconstructed images lose fine details at higher AFs. Accordingly, our objective is to address these limitations and develop a novel CNN for pulmonary gas MRI reconstruction by considering the properties of both k-space and image-domain data and to further improve the reconstruction performance and AF.

In this paper, we propose an encoding enhanced (termed EN$^2$) complex CNN to reconstruct pulmonary gas MR images from highly undersampled k-space data. Particularly, we devise an EN$^2$ convolution kernel for k-space learning, capitalizing on the encoding correlation and integrity within a row or column of k-space. As illustrated in Fig. 1, a row of k-space is acquired in the same frequency-encoding gradient and a column is acquired in the same phase-encoding gradient field [29], [30]. The frequency increment between two rows is Δω as determined by the pulse sequence [31]. The phase increment between two columns is $2\pi/N_x$, where $N_x$ refers to the number of columns in k-space [32]. Thus, there is a strong encoding relationship between data in the same row or column of k-space. EN$^2$ convolution kernel is designed to extract features either row-wise or column-wise. This way of feature extraction is more in line with the properties of k-space data than conventional square convolution kernel and may enable more efficient k-space learning.

Additionally, inspired by the success of the complex convolution strategy for $^1$H MRI reconstruction [33-35], we adopt the complex convolution strategy to learn both magnitude and phase information of pulmonary gas MRI data. Furthermore, we introduce a feature-strengthened modularized unit (FMU) that combines the strengths of residual and dense connections to improve the information flow in the network while enabling effective reuse of feature maps from the preceding layers.

Our main contributions are summarized as follows:

1) EN$^2$ convolution kernel is proposed to maximize the capability of feature learning along k-space encoding direction by resembling the physics of k-space sampling.

2) Complex-valued convolution is used to exploit both magnitude and phase information of pulmonary gas MRI data, improving the data utilization.

3) A feature-strengthened modularized unit is developed to enhance the feature expression ability and further improve the reconstruction performance.

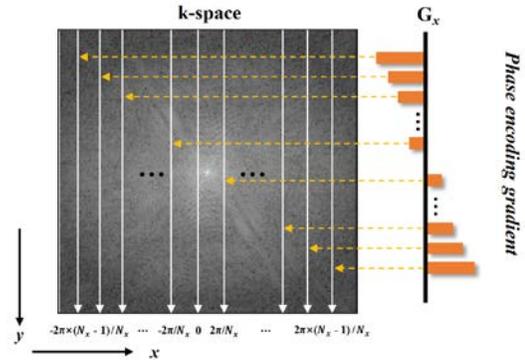

Fig. 1. Schematic diagram of k-space filling under Cartesian sampling.

## II. METHOD

### A. Problem Formulation

Let $\mathbf{x} \in \mathbb{C}^Z$ be a 2D fully sampled pulmonary gas MR image, $\mathbf{y} \in \mathbb{C}^Z$ be the 2D pulmonary gas MR k-space data, $\mathbf{u} \in \mathbb{R}^Z$ denotes a binary undersampling mask. The undersampled k-space data $\mathbf{y}_u$ can be represented as follows:

$$\mathbf{y}_u = \mathbf{u} \circ \mathbf{y} = \mathbf{u} \circ f_{2D}(\mathbf{x}) \quad (1)$$

where $Z = N_y \times N_x$, $N_y$ and $N_x$ represent the number of rows and columns of $\mathbf{y}$, respectively, $\mathbb{C}$ and $\mathbb{R}$ represent the complex and real-valued vector space, respectively, ○ denotes element-wise multiplication, and $f_{2D}$ represents 2D Fourier transform. The reconstruction task is to recover the fully sampled pulmonary MR images from the undersampled k-space data $\mathbf{y}_u$. However, this is an ill-posed problem and only approximate solutions can be obtained [36].

Deep learning can be used for MRI reconstruction by learning the mapping between the undersampled k-space data and the fully sampled images using a convolutional neural network [33] by minimizing:

$$\underset{\theta}{\operatorname{argmin}} L(\mathbf{x}, H(\mathbf{y}_u; \theta)) \quad (2)$$

where $H$ denotes the CNN model that predicts the fully sampled images from the undersampled k-space data, $\theta$ represents the model parameters, and $L$ denotes the loss function.

### B. Proposed Encoding Enhanced Complex CNN

We propose an encoding enhanced (EN$^2$) complex CNN to address some of the challenges in existing DL reconstruction methods for pulmonary gas MRI reconstruction. Fig. 2 provides an overview of the proposed network.

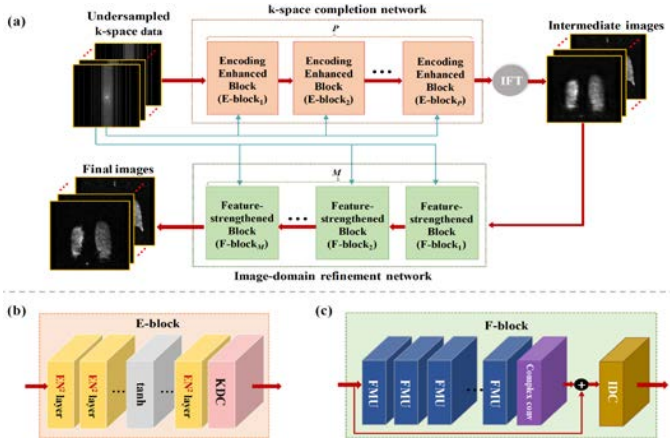

Fig. 2. Architecture of the proposed encoding enhanced complex CNN. (a) An overview of the proposed network that includes a k-space completion network and an image-domain refinement network. The k-space completion network leverages E-blocks to predict fully sampled k-space data and generates the intermediate images through inverse Fourier transform. The image-domain refinement network uses F-blocks to further improve the quality of intermediate images. Complex convolution is adopted in all convolution layers. (b) - (c) The detailed structures of the E-block and F-block.

EN$^2$ complex CNN comprises a k-space completion network and an image-domain refinement network. EN$^2$ complex CNN is featured with EN$^2$ convolution kernel, complex convolution strategy, and a feature-strengthened modularized unit (FMU). EN$^2$ convolution kernel is delicately designed by considering the k-space encoding process to improve learning efficiency. Complex convolution strategy is adopted in both the k-space learning and image-domain learning to maximize the extraction of complex information from the limited MRI data. Note that all activation layers in the proposed network activate real and imaginary channels separately, rather than performing a complex-valued activation operation. FMU is equipped with a novel scheme for information flow by combining residual and dense connections to strengthen the feature expression ability of the network.

The k-space completion network consists of $P$ Encoding Enhanced Blocks (E-blocks) designed for k-space feature extraction. Each E-block contains $Q$ layers with EN$^2$ convolution kernels (termed EN$^2$ layers), a hyperbolic tangent (Tanh) activation layer, and a k-space data consistency (KDC) layer. Tanh activation layer is placed in the middle of the EN$^2$ layers in each E-block. Tanh activation layer was selected for the proposed network after conducting a comparative experiment with other commonly used activation functions, such as ReLU and Leaky ReLU, due to its superior performance. The KDC layer is utilized to enforce the consistency between predicted k-space data and $\mathbf{y}_u$ at the sampled locations [37]. The output of k-space completion network is used to generate intermediate reconstructed images using inverse Fourier transform.

The image-domain refinement network is composed of $M$ Feature-strengthened Blocks (F-blocks). Each F-block contains $R$ consecutive FMUs, followed by a complex convolution layer with a filter size of $3 \times 3$ and an image-domain data consistency (IDC) layer. The input of each F-block and the output of the last complex convolution layer are skip-connected and entered into the IDC. The IDC layer is used to enforce the consistency between the k-space data of the output images and the input undersampled k-space data $\mathbf{y}_u$ at the sampled locations [38].

*1) Encoding enhanced (EN$^2$) convolution kernel*

Many deep learning-based MRI reconstruction methods employ square convolution for both k-space and image-domain learning [39], [40]. In k-space learning, a square convolution kernel maps all the data within a square window to a point in the output feature map. However, the spatial correlation of the k-space data within a square window may not be as strong as that in the image domain [30]. Therefore, the commonly used square convolution kernel may limit the extraction of highly condensed features of k-space data, leading to lower learning efficiency of k-space than image domain. In recognition of this important issue, we propose EN$^2$ convolution kernel by exploring the k-space encoding strategy to more effectively and efficiently extract the complex features of k-space data.

Fig. 3 illustrates the EN$^2$ convolution. An EN$^2$ convolution kernel maps k-space data with the same encoding gradient (i.e., a row or a column of k-space data) to a feature point of the output feature map. We denote F-EN$^2$ and P-EN$^2$ as the EN$^2$ convolution along the frequency-encoding and phase-encoding directions, respectively. In k-space, the relationships between encoding gradient field ($G_x$, $G_y$) and the position in k-space ($k_x$, $k_y$) are given by [31]:

$$k_x(T) = \gamma \int_0^T G_x(t)dt, \text{ and } k_y(T) = \gamma \int_0^T G_y(t)dt \quad (3)$$

where $\gamma$ is the gyromagnetic ratio, $G_x(t)$ and $G_y(t)$ represent the strength of phase-encoding ($x$-direction) and frequency-encoding gradients ($y$-direction) as a function of time, respectively, and $T$ is the time to collect the data sample. Equation (3) demonstrates that k-space data within the same row possess the same frequency-encoding gradient and contain all signals under a frequency encoding. Using this prior, F-EN$^2$ convolution is designed to extract k-space features by rows along frequency-encoding direction, as shown in Fig. 3(a), which can utilize the complete information under a frequency encoding in each convolution. Similarly, P-EN$^2$ convolution is developed to extract k-space features by columns along phase encoding direction, as shown in Fig. 3(c), which can leverage the full information under a phase encoding in each convolution.

F-EN$^2$ or P-EN$^2$ convolution is formulated as follows:

$$\mathbf{F}^v(i) = \mathbf{K}(i) * \mathbf{E} \quad (4)$$

where $\mathbf{E}$ denotes an EN$^2$ convolution kernel of size $1 \times N_x$ when F-EN$^2$ and of size $N_y \times 1$ in cases of P-EN$^2$. $\mathbf{F}^v$ denotes the output feature vector, $\mathbf{K}$ represents the input k-space data, and $i$ is the row index for F-EN$^2$ or column index for P-EN$^2$.

Fig. 3(b) and 3(d) illustrate the process of generating EN$^2$ layer outputs using $N_F$ F-EN$^2$ convolution kernels and $N_P$ P-EN$^2$ convolution kernels, respectively, through convolution and concatenation. Of note, the appropriate form of EN$^2$ convolution kernel (F-EN$^2$ or P-EN$^2$) will be determined based on the characteristics of the k-space dataset. Within a reconstruction network, either F-EN$^2$ or P-EN$^2$ is selected for k-space learning rather than using both sequentially. This is different from separate convolution that decomposes a high-dimension convolution into several low-dimension convolutions in series [34], [41].

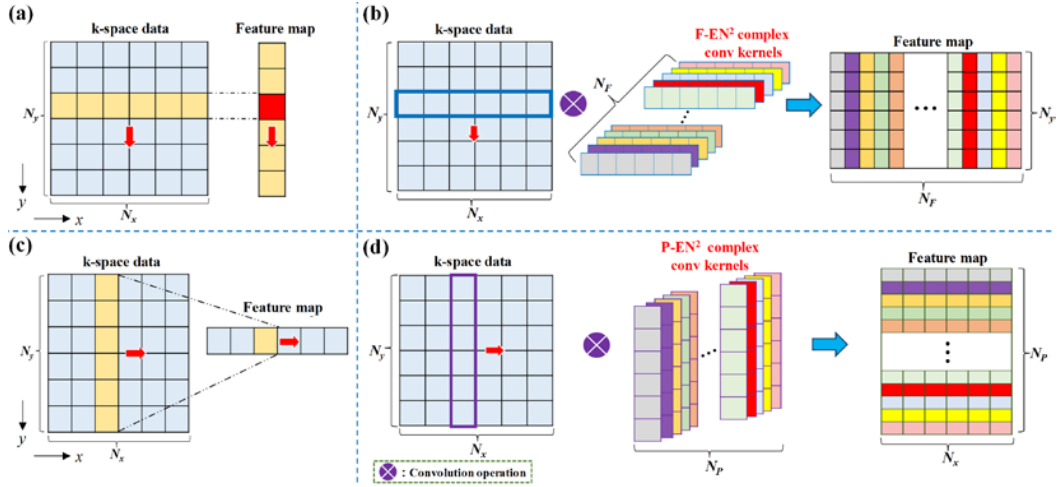

Fig. 3. Schematic of the encoding enhanced (EN$^2$) convolution. (a) Frequency-encoding enhanced (F-EN$^2$) convolution kernel with size $1 \times N_x$. (b) Mapping of an EN$^2$ layer using $N_F$ F-EN$^2$ convolution kernels. (c) Phase-encoding enhanced (P-EN$^2$) convolution kernel with size $N_y \times 1$. (d) Mapping of an EN$^2$ layer using $N_P$ P-EN$^2$ convolution kernels.

*2) Complex convolution strategy*

Complex convolution strategy is adopted in the proposed network to enhance the feature extraction power and data utilization efficiency [34]. Let the complex-valued input k-space data be $C = X + iY$, where $X$ and $Y$ represent the real and imaginary components of $C$, respectively. In a complex convolution layer, $C$ is convolved with a complex-valued filter $W = A + iB$, where $A$ and $B$ denote the real and imaginary parts of the filter. The complex convolution is performed as follows:

$$C * W = (X + iY) * (A + iB) = (X * A - Y * B) + i(X * B + Y * A) \quad (5)$$

where $*$ denotes the convolution operation. Therefore, the complex convolution can be implemented using four real-valued convolutions that explore the relationships between real and imaginary parts of k-space data and complex-valued filters.

*3) Feature-strengthened modularized unit (FMU)*

To strengthen feature propagation and alleviate the gradient-vanishing problem, we design a feature-strengthened modularized unit (FMU) for image-domain learning. The architecture of FMU is shown in Fig. 4.

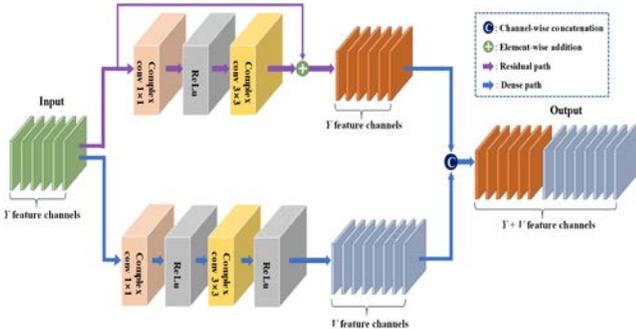

Fig. 4. Structure of the proposed feature-strengthened modularized unit (FMU) in the image-domain refinement network. FMU comprises a residual path (purple) and a dense path (blue). The residual path modifies the feature map through additive operation and the dense path adds new feature channels to the modified feature map through concatenation.

FMU employs residual and dense connections. Residual connection can reuse features from previous layers with low redundancy and dense connection can improve the flow of information between the layers [42], [43]. Unfortunately, residual connection may cause information flow obstruction and dense connection is limited by feature redundancy [44], [45]. The proposed FMU explores the strengths of both residual and dense connections and provides a way to address the limitations of each type of connection when used alone. In an FMU, the input feature maps with $Y$ channels are fed into two paths: the residual path and the dense path. The residual path (purple arrows in Fig. 4) performs residual connection, whereby the feature map is modified through an additive operation. The dense path (blue arrows in Fig. 4) performs dense connection, whereby $V$ new feature channels are added to the modified feature map through concatenation operation. The output of an FMU has $Y + V$ channels and is formulated as:

$$S_{out} = (H_a(S_{in}) \oplus S_{in}) \,||\, H_c(S_{in}) \quad (6)$$

where $S_{in}$ and $S_{out}$ denote the input and output of the FMU, respectively. $H_a$ and $H_c$ refer to the residual path and the dense path, respectively. $H_a$ and $H_c$ employ complex convolution layers with filter size of $1 \times 1$ and $3 \times 3$, respectively, and rectified linear unit (ReLu) activation. $\oplus$ refers to element-wise addition and $||$ indicates channel-wise concatenation. We think the proposed FMU can reuse the features with low redundancy and enable better information flow in the network.

*C. Loss Function*

The loss function of the proposed network $L_s$ is the sum of the loss function of the k-space learning and the loss function of the image-domain learning which are optimized with equal weights. We use $L_w$ loss that is a weighted sum of $L_1$ and $L_2$ loss for both k-space learning and image-domain learning. $L_1$ loss function can prevent over-smoothing and $L_2$ loss function can ensure robust convergence [46]. $L_w$ is defined as:

$$L_w = \alpha L_1 + L_2 \quad (7)$$

where $\alpha$ is a hyper-parameter that weighs the contribution of $L_1$ loss. Hence, the total loss function $L_s$ is formulated as:

$$L_s = L_w(\mathbf{y}, \mathbf{k}_{rec}) + L_w(\mathbf{x}, \mathbf{I}_{rec}) \quad (8)$$

where $\mathbf{k}_{rec}$ denotes the k-space data predicted by the k-space completion network. $\mathbf{I}_{rec}$ denotes the final reconstructed pulmonary gas MR image.

### D. Implementation Details

The k-space completion network utilized $P = 1$ E-block, each comprising $Q = 5$ $EN^2$ layers; the number of $EN^2$ convolution kernels in each $EN^2$ layer was $N_x$ for F-$EN^2$ or $N_y$ for P-$EN^2$. No padding operations were performed within each $EN^2$ layer. For the image-domain refinement network, we used $M = 15$ F-blocks, each encompassing $R = 5$ FMUs.

The weight for the $L_1$ loss function $\alpha$ is set to 1 due to the comparable values of $L_1$ loss and $L_2$ loss for our datasets. Assigning equal weights to the two losses results in a balanced optimization. Our experiments demonstrate that a smaller or larger $\alpha$ will degrade the performance of the proposed network.

We implemented the proposed network using Tensorflow2.0 on a workstation with three Nvidia GeForce RTX 2080 Ti graphics-processing units (GPUs) and an Intel (R) Xeon (R) Gold 6128 central processing unit (CPU). Adam optimizer with an initial learning rate of 0.001, which gradually decayed to $10^{-5}$, was used to train the network. Batch size was 10 and the training was stopped after 200 epochs.

## III. EXPERIMENTAL RESULTS

### A. Datasets

Eighty-five subjects in total were enrolled in this study, including 43 healthy volunteers and 42 patients with lung disease such as chronic obstructive pulmonary disease (COPD), asthma, and so on. All subjects provided informed consents and the experiments were approved by the local Institutional Review Board (IRB). MR imaging was performed using a 1.5T whole-body MRI scanner (Avanto, Siemens Medical Solutions). Enriched $^{129}$Xe gas including 1% enriched xenon (86% 129-isotope), 89% $^4$He, and 10% $N_2$ was polarized using a commercial xenon polarizer (verImagin Healthcare, Wuhan) based on the Rb-$^{129}$Xe spin-exchange optical pumping (Rb-$^{129}$Xe SEOP) method. Then 500 mL hyperpolarized $^{129}$Xe was thawed into a Tedlar bag and mixed with 500 mL medical-grade $N_2$ gas to generate a 1.0 L gas mixture. The available polarization in the bag was approximately 25%. All subjects inhaled the gas mixture from functional residual capacity and then held their breath for data acquisition.

The parameters of pulmonary $^{129}$Xe imaging were as follows: 3D bSSFP sequence, repetition time/echo time (TR/TE) = 4.2/1.9 ms, matrix size = 96 × 84, number of slices = 24, slice thickness = 8 mm, flip angle (FA) = 10°, scan time = 8.4 s. The parameters of $^1$H imaging were as follows: 3D FLASH sequence, TR/TE=2.4/0.7 ms, matrix size = 96 × 84, number of slices = 24, slice thickness = 8 mm, FA = 5°, scan time = 2 s. Anatomic $^1$H images were acquired after functional $^{129}$Xe images during the same breath-hold. The total acquisition time was approximately 11 s. $^{129}$Xe and $^1$H MRI were fully sampled k-space and were single-coil data. The phase encoding direction for both $^{129}$Xe and $^1$H MRI corresponds to the $x$-axis, as illustrated in Fig. 1. The ground-truth images were obtained by directly performing Fourier transform on the fully-sampled k-space data. Before our experiments, $^{129}$Xe and $^1$H MR images were aligned through affine registration.

Note that SNR of the hyperpolarized $^{129}$Xe MRI is relatively low because the longitudinal magnetization of hyperpolarized noble gases decays over time. Thus, it is difficult to acquire hyperpolarized MRI with high SNR and spatial resolution simultaneously [7]. Moreover, the magnetic susceptibility artifacts at the air-tissue interfaces and motion artifacts because of the respiration and cardiovascular pulsation potentially distort and degenerate hyperpolarized MR images. To confirm the image quality, $^{129}$Xe images with signal-to-noise ratio (SNR) less than 6.6 were excluded for downstream analyses. Since noise in MR images follows Rician distribution, SNR is calculated by [47]：

$$\text{SNR} = \frac{\text{Mean}_{\text{signal}} - \text{Mean}_{\text{noise}}}{\text{Std}_{\text{noise}}} \times \sqrt{2 - \frac{\pi}{2}} \quad (9)$$

where $\text{Mean}_{\text{signal}}$ is the mean signal intensity within the lung region, $\text{Mean}_{\text{noise}}$ and $\text{Std}_{\text{noise}}$ are the mean value and standard deviation of the background signal intensities outside the lung.

In this way, a total of 929 $^{129}$Xe slices were included. To facilitate fair performance comparison of different network architectures, the original images of $^{129}$Xe and $^1$H MRI were padded to 96 × 96 to be compatible with all architectures, including U-Net. Each image was normalized to [0, 1]. We randomly selected 68 subjects (700 slices) for training, 8 subjects (101 slices) for validation, and another 9 subjects (128 slices) for testing. Both training and testing sets contained images of healthy volunteers and patients. There is no data overlap between the training and testing sets. A Cartesian scheme was used to obtain undersampled k-space data that corresponded to acceleration factor (AF) of 4 and 6. Specifically, the fully sampled k-space data was retrospectively undersampled in a phase-encoding direction ($x$-direction) according to a variable-density Cartesian random undersampling matrix [39]. The non-acquired k-space data were initialized with zeros to serve as input to the proposed network.

### B. Competing Methods and Assessment Measures
#### 1) Competing methods

We compared our method with a conventional undersampled image reconstruction method SPIRiT [48] and three representative CNN-based MR reconstruction methods: PCNN [22], CasNet [24], and KIKI-net [27]. KIKI-net is a cross-domain reconstruction network, which is trained alternately on k-space and image domain to exploit the information of the raw k-space data. CasNet is the first method using deep learning for pulmonary gas MRI reconstruction. CasNet adopts a cascaded convolution network and incorporates prior knowledge in $^1$H images to refine structural details of the lung. PCNN is a complex CNN for MRI reconstruction, which incorporates a perceptual loss term to suppress incoherent image details. CasNet and PCNN are trained on the image domain alone.

#### 2) Image quality evaluation metrics

We adopted peak signal-to-noise ratio (PSNR) and structure similarity (SSIM) to evaluate the image reconstruction quality. PSNR is measured by mean squared error (MSE), which is the ratio between the peak signal intensity and the intensity of corrupting noise [49]. SSIM is a metric used to evaluate the structural similarity between two images by considering the mean and variance of the image signal intensities [50]. Note that PSNR and SSIM were computed only for the lung region defined by a mask segmented from the corresponding $^1$H images [24].

*3) Assessment via lung function parameters*

To evaluate the potential of the reconstructed images for the assessment of lung function in clinical care, we segmented ventilation defects in the reconstructed pulmonary gas MR images and calculated the ventilation defect percentage (VDP). $^{129}$Xe MRI ventilation defect can be used to quantify pulmonary ventilation distributions and is defined as lung regions with signal intensity below a certain threshold [51]. VDP is one of the commonly used metrics for ventilation defect quantification and represents the ratio of the ventilation defect volume in pulmonary gas MRI to the thoracic volume in $^1$H MRI [52]. The thoracic mask was obtained using seeded region growing followed by manual correction [53]. We employed hierarchical K-means clustering to classify $^{129}$Xe MRI signal intensities to obtain ventilation defect regions [54].

In addition, Dice score was used to evaluate the spatial agreement of the ventilation defects in the fully sampled and the reconstructed $^{129}$Xe MR images [55].

*C. Kernel Selection*

We implemented the proposed algorithm framework in Fig. 2 using F-EN$^2$ convolution kernel sized $1 \times 96$ (Model_F) and P-EN$^2$ convolution kernel sized $96 \times 1$ (Model_P) to determine the suitability of F-EN$^2$ and P-EN$^2$ for our dataset. Of note, Model_F and Model_P were trained, validated, and tested using the same datasets in Section 4.1.

Table I shows that Model_F yielded higher PSNR and SSIM than Model_P at each AF. This suggests that F-EN$^2$ convolution kernel is more suitable for the dataset used in this study. Therefore, we selected F-EN$^2$ convolution in the proposed network for all the other investigations in this work.

TABLE I
QUANTITATIVE RESULTS (PSNR (dB) AND SSIM (%)) OF DIFFERENT MODELS AT VARIOUS AFS [MEAN ± STANDARD DEVIATION].

| AF | Metrics | Model_P | Model_F |
|---|---|---|---|
| 4 | PSNR | 29.98 ± 2.78 | **30.72 ± 2.72** |
|   | SSIM | 86.45 ± 4.70 | **87.51 ± 4.54** |
| 6 | PSNR | 27.44 ± 2.67 | **27.99 ± 2.70** |
|   | SSIM | 77.82 ± 6.71 | **79.22 ± 6.57** |

*Note:* The highest PSNR/SSIM values are bold faced.

*D. Ablation Study*

In this section, we performed a series of ablation studies at an AF of 4 to examine the effects of EN$^2$ convolution kernel, complex-valued convolution, and FMU in the proposed approach. To ensure a fair comparison, we constructed these ablation models with similar numbers of parameters as the proposed network by adjusting the feature channels.

*1) Comparisons of different convolution kernels*

To evaluate the effectiveness of the proposed EN$^2$ convolution kernel, we replaced it with different types of convolution kernels, including square convolution kernels, rectangle convolution kernel, and dilation convolution kernel, while keeping the other components intact.

Fig. 5 shows that EN$^2$ convolution kernel outperformed other compared convolution kernels, highlighting the effectiveness of developing customized convolution kernels based on k-space attributes to enhance reconstruction performance. Notably, for the square convolution kernel, the highest PSNR and SSIM were achieved using the kernel sized $3 \times 3$, while the lowest metrics were generated using the kernel sized $11 \times 11$. The rectangle convolution kernel of size $3 \times 5$ has comparable performance with the square convolution kernel of size $3 \times 3$. Moreover, the performance of the dilated convolution with kernel size of $3 \times 3$ and dilation rate of 2 is inferior to that of the square kernel sized $3 \times 3$.

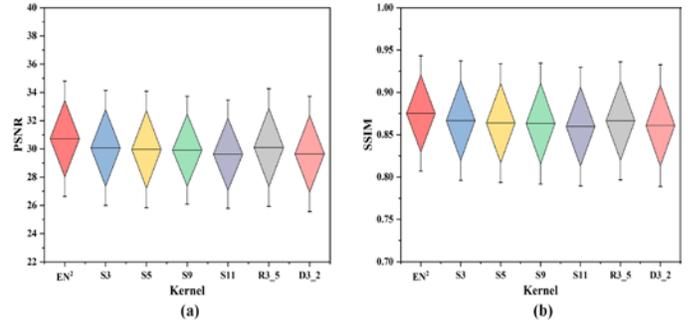

Fig.5. Quantitative comparisons of reconstructions (PSNR and SSIM) using different convolution kernels. S$n$ ($n$ = 3, 5, 9, 11) denotes a square convolution kernel of size $n \times n$. R3_5 denotes a rectangle convolution kernel of size $3 \times 5$. D3_2 denotes a dilated convolution with kernel size of $3 \times 3$ and dilation rate of 2.

*2) Effect of the size of EN$^2$ convolution kernel*

To investigate the impact of incorporating different numbers of k-space data points ($N$) from the same frequency encoding in each convolution, we compared the performance of EN$^2$ convolution with different kernel sizes ($1 \times N$).

Fig. 6 shows that the PSNR and SSIM gradually increased with greater $N$ and the optimal EN$^2$ reconstruction performance was achieved with kernel sized $1 \times 96$. This demonstrates the importance of incorporating the complete set of k-space data from the same encoding in each convolution, which allows for comprehensive utilization of the encoding information.

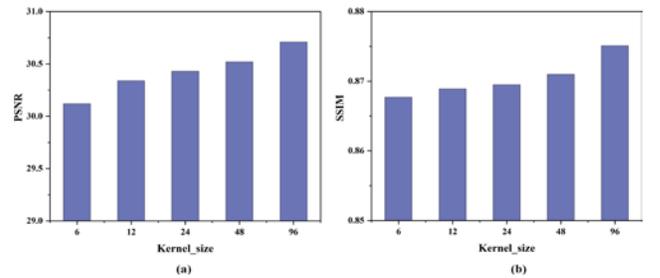

Fig. 6. Quantitative comparisons between the EN$^2$ convolution with different kernel sizes ($1 \times N$, where $N$ = 6, 12, 24, 48, 96).

*3) Comparisons with joint use of F-EN$^2$ and P-EN$^2$*

To investigate whether the combined application of F-EN$^2$ and P-EN$^2$ could further enhance the reconstruction performance, we established two derivative models. The first model sequentially employed F-EN$^2$ and P-EN$^2$ layers (termed EN$^2$_seq). The second model applied both layers in parallel, then merged the generated feature maps using an average operation (termed EN$^2$_merged). Fig. 7 shows that the proposed model using only F-EN$^2$ achieved better reconstruction performance than $EN^2$_seq and $EN^2$_merged. This suggests that the joint use of F-$EN^2$ and P-$EN^2$ may not necessarily further improve performance.

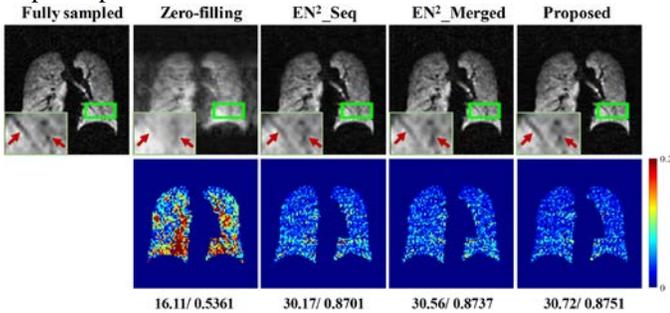

Fig. 7. Comparison between the proposed method and joint use of F-$EN^2$ and P-$EN^2$ layers at an AF of 4. Magnified regions of green boxes are placed at the left bottom of the corresponding images. The error maps show the differences in the lung regions between the reconstructed images and fully sampled images. Average PSNR / SSIM of the entire testing dataset are listed under each reconstruction.

*4) Impact of complex-valued convolution*

To demonstrate the influence of complex-valued convolution on the reconstruction, we replaced the complex-valued convolution with real-valued convolution in the proposed algorithm (termed Proposed_real). Fig. 8 displays the reconstruction results obtained by the proposed method and Proposed_real for a healthy subject and a patient with chronic pulmonary inflammation. The results show that the proposed method reduced pixel-wise reconstruction error and produced sharper boundaries compared to Proposed_real. This substantiates the efficacy of complex-valued convolution in enhancing $^{129}$Xe MRI reconstruction performance.

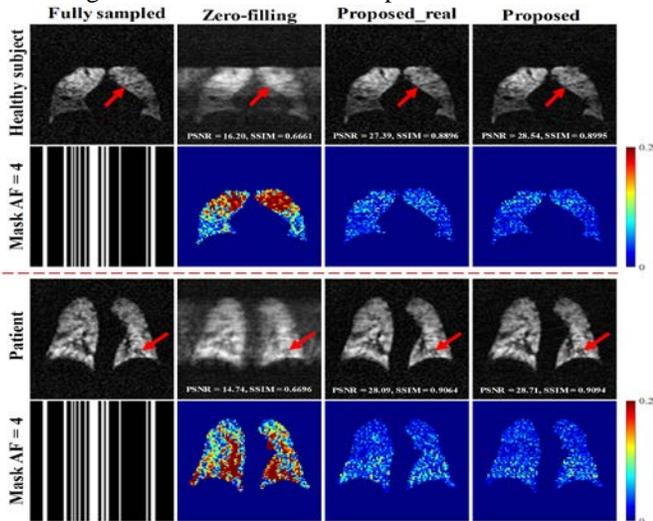

Fig. 8. Comparison between the proposed method and Proposed_real at an AF of 4 (average case). The error maps show the differences in the lung regions between the reconstructed images and fully sampled images.

*5) Strength of the FMU*

To understand the strength of the proposed FMU, we performed three ablation studies. In the first study, we replaced the F-blocks in the proposed network with conventional cascaded convolution blocks (Model_1). In the second study, we removed the dense path of FMU (Model_2). In the third study, we removed the residual path of FMU (Model_3).

Table II shows that Model_1 exhibited the worst reconstruction performance. Model_3 that only includes the dense path of FMU outperformed Model_2 that solely includes the residual path of FMU. This means that both dense and residual connections have a positive impact on the reconstruction performance, with a more pronounced influence from dense connections. These observations align with previous studies [44], [45]. Moreover, the proposed FMU that combines residual and dense connections achieves the best performance, demonstrating the effectiveness of combining these two types of connections for improved results.

TABLE II
ABLATION STUDIES OF FMU [MEAN ± STANDARD DEVIATION].

| Metrics | Model_1 | Model_2 | Model_3 | Proposed |
|---|---|---|---|---|
| PSNR | 29.47 ± 2.68 | 29.92 ± 2.55 | 30.62 ± 2.69 | **30.72 ± 2.72** |
| SSIM (%) | 85.43 ± 4.89 | 85.92 ± 5.01 | 87.32 ± 4.54 | **87.51 ± 4.54** |

*Note:* The highest PSNR/SSIM values are bold faced.

*6) The versatility of $EN^2$ convolution kernel and FMU*

To demonstrate the versatility of the proposed $EN^2$ convolution kernel and FMU, we applied them in the architecture of KIKI-net (termed KIKI-net-E and KIKI-net-F, respectively). These models were built with a similar number of parameters as the original KIKI-net for fair comparison. Table III shows that both the $EN^2$ convolution kernel and FMU enhanced the image reconstruction performance, compared to the original version of KIKI-net. This further confirms the efficacy of $EN^2$ convolution kernel and FMU in improving reconstruction performance and demonstrates their compatibility with various state-of-the-art reconstruction architectures.

TABLE III
COMPARISON OF KIKI-NET WITH AND WITHOUT $EN^2$ KERNEL OR FMU [MEAN ± STANDARD DEVIATION].

| Method | KIKI-net | KIKI-net-E | KIKI-net-F |
|---|---|---|---|
| PSNR | 29.68 ± 2.99 | **30.06 ± 2.95** | 29.99 ± 2.73 |
| SSIM (%) | 86.80 ± 4.65 | **87.24 ± 4.32** | 87.06 ± 4.50 |

*Note*: The highest PSNR and SSIM are bold faced.

*E. Comparisons with State-of-the-Arts*

*1) Image reconstruction performance*

Table IV compares the image reconstruction performance of the proposed approach and SPIRiT, CasNet, PCNN, and KIKI-net at different AFs in different subject groups. These results demonstrate that the proposed method consistently performed better than the other methods for various clinical indications at each AF.

Compared with the KIKI-net, the proposed network reconstructed $^{129}$Xe MR images with PSNR 3.50% and 4.91% higher at AF of 4 and 6 for the entire dataset, respectively. Of note, the cross-domain learning-based methods (KIKI-net and the proposed method) outperformed the image-domain learning-based methods (CasNet and PCNN), consistent with previous findings [27], [40]. These deep learning-based methods outperformed SPIRiT by a large margin. Table V presents a comparison of model parameters among deep learning-based methods. The results demonstrate that the proposed method and KIKI-net have comparable model parameters, which are significantly fewer than those of CasNet and PCNN.

TABLE IV
QUANTITATIVE RESULTS OF THE COMPARISON METHODS AT DIFFERENT AFS [MEAN ± STANDARD DEVIATION].

| Clinical indications | Model | AF = 4 | | AF = 6 | |
|---|---|---|---|---|---|
| | | PSNR | SSIM (%) | PSNR | SSIM (%) |
| Healthy | Zero-filling | 16.11 ± 2.14 | 53.61 ± 4.98 | 15.10 ± 2.14 | 45.84 ± 5.41 |
| | SPIRiT | 22.26 ± 2.17 | 67.12 ± 6.09 | 20.67 ± 2.16 | 60.49 ± 6.04 |
| | CasNet | 25.52 ± 1.87 | 80.11 ± 3.94 | 23.40 ± 2.01 | 72.66 ± 5.05 |
| | PCNN | 27.65 ± 1.56 | 85.12 ± 3.67 | 25.10 ± 1.26 | 75.61 ± 5.30 |
| | KIKI-net | 28.02 ± 2.11 | 87.23 ± 3.44 | 25.53 ± 1.88 | 79.15 ± 5.09 |
| | Proposed | **29.20 ± 1.37** | **87.67 ± 3.44** | **26.65 ± 1.26** | **79.47 ± 5.15** |
| Patient | Zero-filling | 17.32 ± 2.16 | 53.42 ± 8.66 | 16.26 ± 2.01 | 46.19 ± 8.81 |
| | SPIRiT | 25.27 ± 2.97 | 67.68 ± 9.90 | 23.89 ± 2.73 | 61.84 ± 9.50 |
| | CasNet | 27.19 ± 3.05 | 78.14 ± 7.21 | 25.35 ± 2.86 | 70.92 ± 8.60 |
| | PCNN | 29.53 ± 2.86 | 84.24 ± 5.60 | 27.16 ± 2.58 | 74.96 ± 7.52 |
| | KIKI-net | 30.38 ± 3.04 | 86.62 ± 5.08 | 27.32 ± 3.20 | 78.57 ± 7.24 |
| | Proposed | **31.36 ± 2.89** | **87.44 ± 4.95** | **28.56 ± 2.94** | **79.11 ± 7.11** |
| Total | Zero-filling | 16.96 ± 2.22 | 53.48 ± 7.83 | 15.91 ± 2.11 | 46.08 ± 7.99 |
| | SPIRiT | 24.37 ± 3.07 | 67.51 ± 8.92 | 22.93 ± 2.96 | 61.44 ± 8.62 |
| | CasNet | 26.69 ± 2.85 | 78.73 ± 6.45 | 24.70 ± 2.78 | 71.44 ± 7.73 |
| | PCNN | 28.97 ± 2.68 | 84.50 ± 5.11 | 26.66 ± 2.39 | 75.16 ± 6.92 |
| | KIKI-net | 29.68 ± 2.99 | 86.80 ± 4.65 | 26.68 ± 3.04 | 78.75 ± 6.67 |
| | Proposed | **30.72 ± 2.72** | **87.51 ± 4.54** | **27.99 ± 2.70** | **79.22 ± 6.57** |

*Note*: The highest PSNR and SSIM are bold faced.

TABLE V
COMPARISON OF THE NUMBER OF TRAINABLE PARAMETERS AMONG DIFFERENT METHODS.

| Methods | CasNet | PCNN | KIKI-net | Proposed |
|---|---|---|---|---|
| Parameters | 94.17 M | 18.24 M | 2.06 M | **1.91 M** |

*Note*: The fewest parameters are bold faced.

Fig. 9 shows that our approach restored $^{129}$Xe MR images with structures in more detail and textures in higher degree of consistency with the fully sampled images, compared with the other methods for a healthy volunteer and a pulmonary tuberculosis patient at an AF of 6; this is consistent with the quantitative results in Table IV. For example, the images reconstructed by CasNet, PCNN, and KIKI-net were blurred owing to the high AF, especially in low ventilation regions (indicated by red arrows). The proposed approach is more accurate with fewer reconstruction errors.

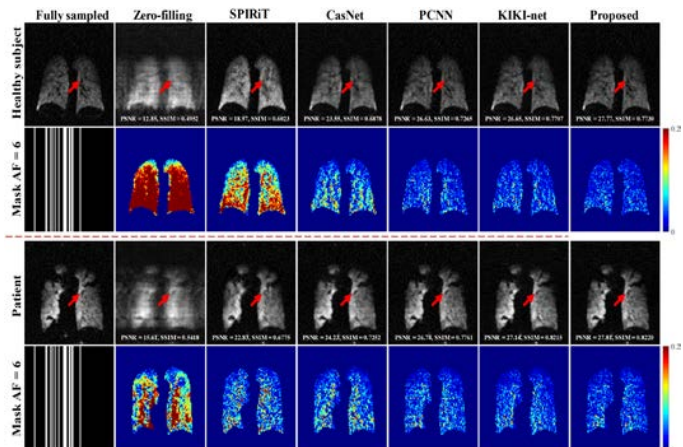

Fig. 9. Reconstruction results of two representative slices (average case) at an AF of 6: one slice from a healthy volunteer and the other from a pulmonary tuberculosis patient. The corresponding error maps show the differences in the lung regions between the reconstructed images and fully sampled images.

*2) Robustness to noise*

Systemic noise has been recognized as one of the major issues in pulmonary $^{129}$Xe MR imaging. The robustness to noise for all the methods was evaluated. In particular, complex-valued Gaussian noises with mean ($\mu$) of 0 and standard deviation ($\sigma$) of 0.01, 0.05, and 0.1 were added to the original k-space data as algorithm inputs. The quantitative reconstruction results for different methods at an AF of 4 are presented in Fig. 10.

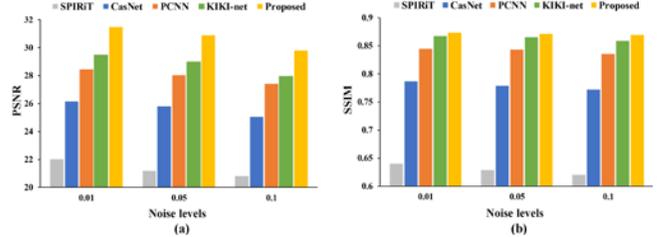

Fig. 10. The histograms of the average values of PSNR and SSIM under three levels of added Gaussian noise ($\mu = 0$, $\sigma = 0.01$, 0.05, and 0.1) for all test data at an AF of 4.

Fig. 10 shows that the proposed method achieved the highest PSNR and SSIM among the comparison methods at all the noise levels. In addition, at the highest noise level ($\sigma = 0.1$), the PSNR of the proposed method was decreased by 4.8% compared with the noise-free case ($\sigma = 0$), and these were 14.56%, 6.1%, 5.4%, and 5.8% for SPIRiT, CasNet, PCNN, and KIKI-net, respectively.

*3) Comparisons of the lung function parameters*

Table VI presents the Dice scores for the ventilation defect regions segmented from the $^{129}$Xe MR images reconstructed using all the comparative methods at different AFs. At each AF, the proposed method achieved the highest mean Dice with the lowest standard deviation for the healthy subject and patient groups and the entire dataset. This means that the proposed method could be used to generate $^{129}$Xe MRI ventilation defect measurements highly consistent with fully sampled images.

TABLE VI
DICE SCORES CALCULATED BETWEEN THE VENTILATION DEFECT REGIONS OF THE RECONSTRUCTED IMAGES AND FULLY SAMPLED IMAGES AT DIFFERENT AFS [MEAN ± STANDARD DEVIATION].

| Clinical indications | Model | AF = 4 | AF = 6 |
|---|---|---|---|
| Healthy | SPIRiT | 0.6489 ± 0.1256 | 0.6921 ± 0.0975 |
| | CasNet | 0.7355 ± 0.0771 | 0.6921 ± 0.0975 |
| | PCNN | 0.7843 ± 0.0600 | 0.7044 ± 0.0787 |
| | KIKI-net | 0.8176 ± 0.0526 | 0.7468 ± 0.0762 |
| | Proposed | **0.8260 ± 0.0451** | **0.7623 ± 0.0708** |
| Patient | SPIRiT | 0.7721 ± 0.1180 | 0.7714 ± 0.1607 |
| | CasNet | 0.8192 ± 0.1102 | 0.7714 ± 0.1607 |
| | PCNN | 0.8563 ± 0.0790 | 0.8094 ± 0.1062 |
| | KIKI-net | 0.8733 ± 0.0684 | 0.8267 ± 0.1143 |
| | Proposed | **0.8812 ± 0.0618** | **0.8363 ± 0.0991** |
| Total | SPIRiT | 0.7380 ± 0.1317 | 0.7494 ± 0.1498 |
| | CasNet | 0.7960 ± 0.1085 | 0.7494 ± 0.1498 |
| | PCNN | 0.8363 ± 0.0807 | 0.7803 ± 0.1097 |
| | KIKI-net | 0.8579 ± 0.0689 | 0.8045 ± 0.1090 |
| | Proposed | **0.8659 ± 0.0626** | **0.8158 ± 0.0976** |

*Note*: The highest Dice scores are bold faced.

Fig. 11 shows the ventilation defect segmented from the results of all the reconstruction methods at different AFs. Consistent with the quantitative results, the segmentation results from the reconstructed images of the proposed method showed better agreement with that from fully sampled images than the other comparison methods. Specifically, at a relatively low AF of 4, the ventilation defect regions obtained by the proposed method were very similar to that generated from the fully sampled images for both the healthy subject and the patient. In contrast, there were visible differences between the ventilation defect regions calculated from the other comparative methods and the fully sampled images, especially for the patient group (see blue arrows). At a higher AF of 6, the ventilation defect regions obtained by the proposed method were more similar to the fully sampled images, compared with the other methods. For example, some small ventilation defect regions in the right lower lobe of the healthy subject were restored by the proposed method whereas the other methods failed as indicated by the blue arrows.

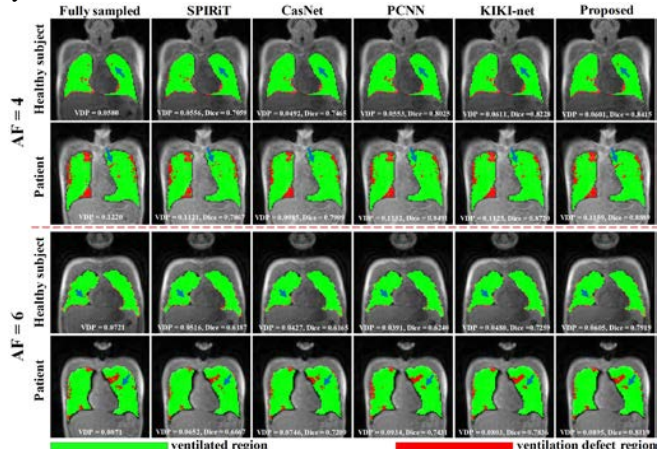

Fig. 11. Segmentation results of representative slices (average case) at different AFs. The green color denotes ventilated regions and the red color denotes the ventilation defect regions. VDP values and Dice scores for each reconstructed image are listed below the segmentation results.

We also evaluated the utility of the proposed approach by comparing the VDP calculated from the gas MR images reconstructed by our method and KIKI-net (providing the highest PSNR and SSIM among the comparative methods) with the VDP derived from the fully sampled images at different AFs. As shown in Fig. 12(a) and 12(b), our approach yielded higher Pearson correlation coefficients ($r$) for VDP than KIKI-net at AF of 4 ($r = 0.978$ for our approach and 0.960 for KIKI-net). Notably, the VDP correlation between our approach and the fully sampled reconstruction ($r = 0.951$) is much higher than that between KIKI-net and the fully sampled images ($r = 0.907$) at AF of 6.

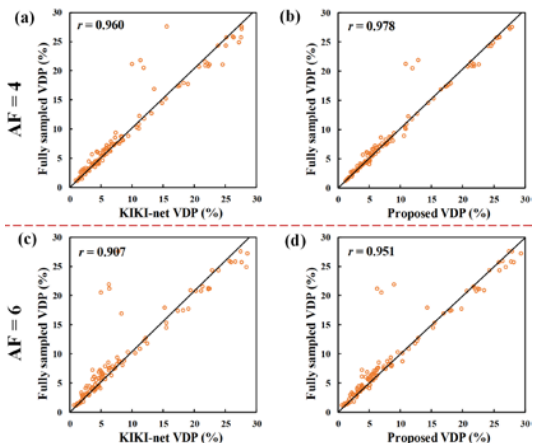

Fig. 12. VDP comparisons between the fully sampled images and reconstructed images of different methods at AFs of 4 and 6. Pearson correlation coefficients ($r$ values) are listed in the scatter plots.

### 4) Performance evaluation on unpadded original dataset

To ensure a fair comparison among various reconstruction methods, the original data were padded to $96 \times 96$. Yet, such padding operations may influence the reconstruction results [56]. To robustly demonstrate the superiority of our proposed method, we compared its performance with KIKI-net on original data (of size $96 \times 84$) without padding at an AF of 4. Table VII shows that the proposed network outperformed KIKI-net regardless of whether data were padded or not, further affirming the effectiveness of our method. However, it should be noted that both our method and KIKI-Net show slight performance variations across different data sizes. Consequently, in practical applications, it is preferable to avoid preprocessing steps like padding that might impact reconstruction performance.

TABLE VII
COMPARISONS OF KIKI-NET AND THE PROPOSED METHOD USING DIFFERENT DATA SIZES AT AN AF OF 4
[MEAN (STANDARD DEVIATION)].

| Methods | $96 \times 84$ | | $96 \times 96$ | |
|---|---|---|---|---|
| | KIKI-net | Proposed | KIKI-net | Proposed |
| PSNR | 29.97 ± 3.16 | **30.79 ± 2.79** | 29.68 ± 2.99 | **30.72 ± 2.72** |
| SSIM (%) | 86.91 ± 4.36 | **87.96 ± 4.35** | 86.80 ± 4.65 | **87.51 ± 4.54** |

*Note*: The highest PSNR and SSIM are bold faced.

### F. Extension to the Lung $^1$H MRI

To evaluate the generality of the proposed method, we compared the reconstruction performance of the proposed methods with SPIRiT, CasNet, PCNN, and KIKI-net on lung $^1$H MRI dataset at an AF of 4. In the $^1$H MRI dataset, there were 85 subjects and each subject had 24 slices. We randomly selected 68 subjects for training, 8 for validation, and 9 for testing. Both training and testing sets contained healthy volunteers and patients. Representative reconstruction results are shown in Fig. 13. The results show that the proposed method restored fine structures in $^1$H MR images with the highest PSNR and SSIM, compared with the other methods. This demonstrates the utility and generality of the proposed approach for use in $^1$H MRI reconstruction.

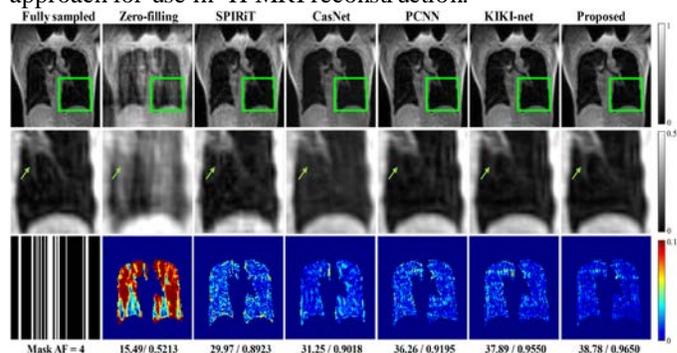

Fig. 13. Reconstruction results of lung $^1$H MRI using different methods at an AF of 4 (average case). The first row shows the reconstructed images. Magnified regions of green boxes are placed on the second row, and the error maps showing the differences in the lung regions are placed on the third row. Average PSNR / SSIM of the entire testing dataset are listed under each reconstruction.

## IV. DISCUSSIONS

In this study, we develop an encoding enhanced (EN$^2$) complex CNN to reconstruct pulmonary MR images from highly undersampled k-space data. In contrast to other CNN-based MRI reconstruction networks that apply square convolution kernels for k-space and image-domain learning [37], [39], we design a novel $EN^2$ convolution kernel to extract the features of k-space along encoding direction based on the encoding properties of k-space data acquisition. This convolution strategy improves the efficiency of k-space learning. The network is implemented with complex convolution to maximize feature extraction capability from the limited gas MRI data. In addition, we propose an FMU module to further strengthen the representation capability of the network. Experiments demonstrate the effectiveness of these elements and the proposed approach.

The size of the $EN^2$ convolution kernel ($1 \times 96$) is larger than the commonly used square convolution kernels (i.e., $3 \times 3$). To investigate the influence of receptive field, we compared the $EN^2$ convolution kernel with square convolution kernel of different sizes (up to $11 \times 11$). Results showed that $EN^2$ convolution kernel achieved the best reconstruction results (see Fig. 5). This indicates that commonly used square convolution using large kernels does not necessarily lead to better results and in fact, we showed that square convolution with kernel size of $3 \times 3$ outperformed the other cases. This suggests square convolution for k-space learning is not optimal and the receptive field of kernels is not a determining factor for the observed performance improvement for the proposed method. The results of rectangle convolution and dilation convolution further support this conclusion.

The improvement derived from $EN^2$ convolution is due to its unique ability to fully leverage the integrity and correlation of encoding information within rows or columns in k-space. This yields a more accurate representation of k-space. Specifically, k-space data with the same encoding gradient (a row or column) have a stronger correlation than that within a square window. In a square patch of k-space, both the phase-encoding and frequency-encoding gradients are different for different data points, leading to potentially weaker encoding correlations. Moreover, the square convolution kernel only scans a limited square region of k-space data in each convolution, resulting in incomplete encoding in each direction. Furthermore, the spatial correlation of k-space data points within a square window is not as strong as that of image domain data, due to each data point in k-space containing information from the entire MR image. Consequently, the advantage of the square convolution kernel in extracting local features is diminished in k-space learning. As for the $EN^2$ convolution kernel (taking F-$EN^2$ for example), the complete information under a frequency encoding is utilized in each convolution. As a result, $EN^2$ convolution kernel can capture highly condensed and non-redundant features of k-space, enabling more effective feature mining than the square convolution kernel. To our knowledge, this is the first investigation of MRI reconstruction using a new form of convolution kernel that considers the characteristics of k-space data, which distinctly differs from existing methods.

Here, we showed that selecting the suitable form of the $EN^2$ convolution kernel (F-$EN^2$ or P-$EN^2$) according to the dataset is helpful in obtaining better reconstruction results. Our dataset includes k-space data that is undersampled in the phase-encoding direction (*x*-direction), meaning that the k-space is either fully sampled or zero-filled in the same column. F-$EN^2$ convolution kernel can exploit the relationship between sampled and unsampled data in each convolution, which is not available to the P-$EN^2$. Therefore, F-$EN^2$ performs better than P-$EN^2$ in our dataset, as shown in Table I. However, it is essential to recognize that the superiority of F-$EN^2$ might not hold across all types of datasets. For example, undersampling scheme maintains frequency continuity within the fully sampled columns but compromises phase continuity in each row. For a smaller k-space matrix (e.g., $64 \times 64$), this disruption of continuity can have a stronger negative impact on k-space learning efficiency, potentially leading to better performance of P-$EN^2$ than F-$EN^2$. Thus, we think that both F-$EN^2$ and P-$EN^2$ can be beneficial and the choice depends on the characteristics of the datasets.

Remarkably, we observed that the joint implementation of F-$EN^2$ and P-$EN^2$ kernels did not enhance the reconstruction performance (Fig. 7). In $EN^2$\_seq, the convolution operations of the F-$EN^2$ disrupt the consistency of phase encoding within the same column data. Given that the P-$EN^2$ kernel is designed on the premise of phase consistency within the same column k-space data, the P-$EN^2$ layer struggles to efficiently extract features from feature maps extracted by the preceding F-$EN^2$ layer. Similarly, the convolution operations of the P-$EN^2$ layer compromise the frequency consistency within the same row data. This negatively impacts subsequent feature extraction by the F-$EN^2$, resulting in reduced reconstruction performance than using the F-$EN^2$ only. Moreover, the result of $EN^2$\_merged demonstrates that adequate exploitation of the correlation between k-space data with the same frequency encoding might be sufficient for effective k-space feature extraction. Given identical parameter sizes, the numbers of F-$EN^2$ and P-$EN^2$ kernels in the $EN^2$\_merged are only half that of the F-$EN^2$ kernels in the proposed model. In our dataset, where F-$EN^2$ outperforms P-$EN^2$, the benefits of incorporating additional encoding direction information may not compensate for the negative impact of less comprehensive feature extraction due to fewer F-$EN^2$ kernels. Additionally, we acknowledge that the currently adopted average merging strategy may not optimally take advantage of combining information from both encoding directions and more comprehensive investigation of the feature fusion methods potentially improves the performance of $EN^2$\_Merged. However, the fusion of feature maps from different encoding directions, especially in the context of k-space, is a challenging task that remains underexplored. We plan to investigate effective k-space feature fusion methods in our future efforts with high priority.

Previous research shows that complex-valued convolution considers the relationship between the real and imaginary components of k-space data, allowing the network to gain greater representation power than real-valued convolution [33]. Our results confirm the effectiveness of complex-valued convolution in the context of pulmonary gas MRI reconstruction (Fig. 8). Complex-valued convolution treats complex-valued k-space data as a whole, which is decomposed into four different real-valued convolutions to extract the hidden information in the real and imaginary parts of k-space. This is critical for generating an image. In contrast, real-valued convolution deals with the real and imaginary parts of k-space separately, leading to difficulties in exploiting the correlation between the two components and suboptimal reconstruction as a consequence. Therefore, we think complex-valued convolution can better deal with learning from the limited pulmonary gas MRI data. Moreover, to further enhance the feature expression capability, we propose an FMU module combining residual and dense connections. In CNN, residual connection enables feature reuse and dense connection enables new feature exploration, and both mechanisms have been widely used for MRI reconstruction [20] [21]. However, residual connection is not optimal for exploring new features and dense connection suffers from high redundancy [44]. FMU integrates the two types of connections and performs residual connection before dense connection. This leads to higher flexibility in learning new features while maintaining low feature redundancy. The results of the ablation study on FMU demonstrate the effectiveness of FMU in improving reconstruction performance.

Our approach outperformed several state-of-the-art reconstruction methods under different AFs quantitatively and qualitatively (see TABLE IV, Fig. 9). In particular, our method can well preserve lung structures and restore most of the texture details at a relatively high AF of 6, which was not available using the other compared methods and was 50% higher than existing lung gas MRI reconstruction methods [24], [25]. We also evaluated our method and KIKI-net (the best performing methods at AFs of 4 and 6) at an AF of 8, whereby the original k-space data (96 × 96) was highly undersampled and contained little information. While we observed substantially reduced reconstruction performance for both our method and KIKI-net as expected, our approach still outperformed KIKI-net with higher quantitative metrics and more details, as shown in Fig. 14. These results indicate that our method is useful for highly undersampled reconstruction. Additionally, our method demonstrated generality for $^1$H lung MRI reconstruction (Fig. 13), which has also shown numerous promises for lung disease research, suggesting the utility of the proposed approach for a wide range of applications.

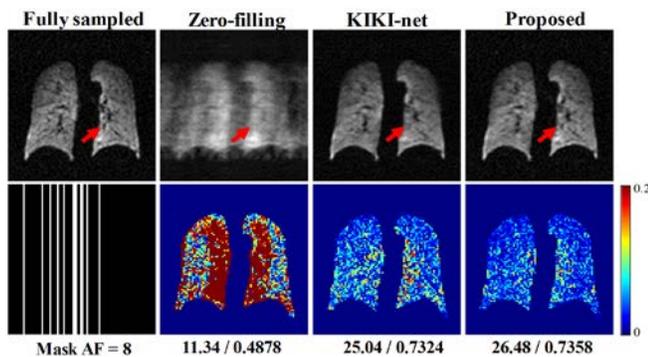

Fig. 14. Reconstruction results using different methods at an AF of 8 (average case). Reconstructed images and error maps are placed on the first and second rows. Average PSNR / SSIM of the entire testing dataset are listed under each reconstruction.

We further demonstrated the clinical utility of the proposed method by segmenting ventilation defects and quantifying VDP from the reconstructed $^{129}$Xe MR images. Our approach demonstrated higher fidelity with the fully sampled images, compared with the other methods (see Table VI and Fig. 11). The Dice scores for ventilation defect segmentation were promising considering the small size of defect regions that account for less than 10% of the lung for most of the cases. The VDP measurements derived from our method had stronger correlations with the fully sampled images (see Fig. 12). All of this is consistent with translating $^{129}$Xe MRI into broad research and clinical patient care.

Whether to include a bias term in the convolutional layer that operates on k-space data remains debated. While this operation may be destructive and shift image signal intensities or can lead to instability during training [17], other studies [27], [29], [57] showed that this term can enhance data fitting capability and flexibility of deep models. The reason behind this is not entirely clear and may be related to the characteristics of training data, architecture of the model, optimization schemes, etc. However, a convolutional layer without a bias term could be viewed as a special case of an ordinary convolution. In this situation, the bias is fixed to zero rather than a learnable parameter. In our approach, we compared $EN^2$ layers with and without a bias term and observed stable training for both cases with PSNR of 30.72 and SSIM of 0.8751 when a bias term was used and PSNR of 30.43 and SSIM of 0.8709 when it was not. Therefore, we opted to incorporate a bias term in the $EN^2$ layers in the proposed approach.

We acknowledge that this study still has limitations. For example, the proposed encoding enhanced complex CNN was only validated on lung MRI data with Cartesian undersampling while other k-space acquisition methods, including spiral, and radial trajectories, are also widely used. Adoption of the proposed method for use in these sampling patterns warrants further investigation.

## V. CONCLUSION

In conclusion, we develop a cross-domain learning based approach for highly undersampled pulmonary $^{129}$Xe and $^1$H MRI reconstruction. Our approach employs a novel convolution scheme by exploiting the k-space data properties, complex convolution to maximize the data use efficiency, and a feature-strengthened unit to enhance feature expression capability. We showed that the proposed approach reconstructed $^{129}$Xe and $^1$H MR images with high accuracy at 6-fold undersampling and yielded lung function measurements that are highly consistent with fully sampled images. As such, our approach may facilitate the integration of deep learning for fast lung MR imaging.

## REFERENCES

[1] A. I. Aviles-Rivero, N. Debroux, G. Williams, M. J. Graves, and C. B. Schönlieb, "Compressed sensing plus motion (CS + M): A new perspective for improving undersampled MR image reconstruction," *Med. Image Anal.*, vol. 68, p. 101933, Feb. 2021.

[2] M. S. Albert et al., "Biological magnetic resonance imaging using laser-

polarized $^{129}$Xe," *Nature*, vol. 370, no. 6486, pp. 199-201, Jul. 1994.

[3] J. M. Reinhardt, K. Ding, K. Cao, G. E. Christensen, E. A. Hoffman, and S. V. Bodas, "Registration-based estimates of local lung tissue expansion compared to xenon CT measures of specific ventilation," *Med. Image Anal.*, vol. 12, no. 6, pp. 752-763, Dec. 2008.

[4] H. Li et al., "Damaged lung gas exchange function of discharged COVID-19 patients detected by hyperpolarized $^{129}$Xe MRI," *Sci. Adv.*, vol. 7, no. 1, p. eabc8180, Jan. 2021.

[5] F. Guo et al., "Globally optimal co-segmentation of three-dimensional pulmonary $^{1}$H and hyper-polarized $^{3}$He MRI with spatial consistence prior," *Med. Image Anal.*, vol. 23, no. 1, pp. 43-55, Jul. 2015.

[6] H. U. Kauczor et al., "Imaging of the lungs using $^{3}$He MRI: Preliminary clinical experience in 18 patients with and without lung disease," *J. Magn. Reson. Imaging*, vol. 7, no. 3, pp. 538-543, May 1997.

[7] S. Xiao et al., "Considering low-rank, sparse and gas-inflow effects constraints for accelerated pulmonary dynamic hyperpolarized $^{129}$Xe MRI," *J. Magn. Reason.*, vol. 290, pp. 29-37, May 2018.

[8] R. F. Lee, G. Johnson, R. L. Grossman, B. Stoeckel, R. Trampel, and G. McGuinness, "Advantages of parallel imaging in conjunction with hyperpolarized helium - A new approach to MRI of the lung," *Magn. Reson. Med.*, vol. 55, no. 5, pp. 1132-1141, May 2006.

[9] M. Salerno, T. A. Altes, J. R. Brookeman, E. E. Lange, and J. P. Mugler, "Dynamic spiral MRI of pulmonary gas flow using hyperpolarized $^{3}$He: preliminary studies in healthy and diseased lungs," *Magn. Reson. Med.* vol. 46, no. 4, pp. 667-677, Oct. 2001.

[10] J. M. Wild et al., "Dynamic radial projection MRI of inhaled hyperpolarized $^{3}$He gas," *Magn. Reson. Med.*, vol. 49, no. 6, pp. 991-997, Jun. 2003.

[11] M. Zhang et al., "High-dimensional embedding network derived prior for compressive sensing MRI reconstruction," *Med. Image Anal.*, vol. 64, p. 101717, Aug. 2020.

[12] S. Ajraoui, K. J. Lee, M. H. Deppe, S. R. Parnell, J. Parra-Robles, and J. M. Wild, "Compressed sensing in hyperpolarized $^{3}$He lung MRI," *Magn. Reson. Med.*, vol. 63, no. 4, pp. 1059-1069, May 2010.

[13] S. Ajraoui, J. Parra-Robles, and J. M. Wild, "Incorporation of prior knowledge in compressed sensing for faster acquisition of hyperpolarized gas images," *Magn. Reson. Med.*, vol. 69, no. 2, pp. 360-369, Feb. 2013.

[14] S. Xiao et al., "Highly and adaptively undersampling pattern for pulmonary hyperpolarized $^{129}$Xe dynamic MRI," *IEEE Trans. Med. Imaging*, vol. 38, no. 5, pp. 1240-1250, May 2019.

[15] Q. Huang et al., "Dynamic MRI reconstruction with end-to-end motion-guided network," *Med. Image Anal.*, vol. 68, p. 101901, Feb. 2021.

[16] Z. Wang et al., "One-dimensional deep low-rank and sparse network for accelerated MRI," *IEEE Trans. Med. Imaging*, vol. 42, no. 1, pp. 79-90, Jan. 2023.

[17] M. Akçakaya, S. Moeller, S. Weingärtner, and K. Uğurbil, "Scan-specific robust artificial-neural-networks for k-space interpolation (RAKI) reconstruction: Database-free deep learning for fast imaging," *Magn. Reson. Med.*, vol. 81, no. 1, pp. 439-453, Jan. 2019.

[18] K. Hammernik, et al., "Learning a variational network for reconstruction of accelerated MRI data," *Magn. Reson. Med.*, vol. 79, no. 6, pp. 3055-3071, Jun. 2018.

[19] S. Wang et al., "Accelerating magnetic resonance imaging via deep learning," in *IEEE 13th International Symposium on Biomedical Imaging (ISBI)*, 2016, pp. 514-517.

[20] A. Aghabiglou and E. M. Eksioglu, "MR image reconstruction using densely connected residual convolutional networks," *Comput. Biol. Med.*, vol. 139, p. 105010, Dec. 2021.

[21] D. Lee, J. Yoo, S. Tak, and J. C. Ye, "Deep residual learning for accelerated MRI using magnitude and phase networks," *IEEE Trans. Biomed. Eng.*, vol. 65, no. 9, pp. 1985-1995, Apr. 2018.

[22] D. Shen et al., "Rapid reconstruction of highly undersampled, non-Cartesian real-time cine k-space data using a perceptual complex neural network (PCNN)," *NMR Biomed.*, vol. 34, no. 1, p. 4405, Jan. 2021.

[23] C. Duan et al., "Accelerating susceptibility-weighted imaging with deep learning by complex-valued convolutional neural network (ComplexNet): validation in clinical brain imaging," *Eur. Radiol.*, vol. 32, pp. 5679-5687, Feb. 2022.

[24] C. Duan et al., "Fast and accurate reconstruction of human lung gas MRI with deep learning," *Magn. Reson. Med.* vol. 82, no. 6, pp. 2273-2285, Dec. 2019.

[25] C. Duan et al., "Accelerate gas diffusion-weighted MRI for lung morphometry with deep learning," *Eur. Radiol.*, vol. 32, no. 1, pp. 702-713, 2022.

[26] T. Du, Y. Zhang, X. Shi, S. Chen, "Multiple Slice k-space Deep Learning for Magnetic Resonance Imaging Reconstruction," in *International Conference of the IEEE Engineering in Medicine & Biology Society (EMBC)*, 2020, pp. 1564-1567.

[27] T. Eo, Y. Jun, T. Kim, J. Jang, H. J. Lee, and D. Hwang, "KIKI-net: cross-domain convolutional neural networks for reconstructing undersampled magnetic resonance images," *Magn. Reson. Med.*, vol. 80, no. 5, pp. 2188-2201, Nov. 2018.

[28] T. Du et al., "Adaptive convolutional neural networks for accelerating magnetic resonance imaging via k-space data interpolation," *Med. Image Anal.*, vol. 72, p. 102098, Aug. 2021.

[29] T. Eo, H. Shin, Y. Jun, T. Kim, and D. Hwang, "Accelerating Cartesian MRI by domain-transform manifold learning in phase-encoding direction," *Med. Image Anal.*, vol. 63, p. 101689, Jul. 2020.

[30] T. A. Spraggins, "A perspective on k-space," *Radiology*, vol. 199, no. 3, pp. 874-875, Jun. 1996.

[31] C. B. Paschal and H. D. Morris, "K-Space in the Clinic," *J. Magn. Reson. Imaging*, vol. 19, no. 2, pp. 145-159, Feb. 2004.

[32] J. Hennig, "K-space sampling strategies," *Eur. Radiol.*, vol. 9, pp. 1020-1031, Jul. 1999.

[33] E. Cole, J. Cheng, J. Pauly, and S. Vasanawala, "Analysis of deep complex-valued convolutional neural networks for MRI reconstruction and phase-focused applications," *Magn. Reson. Med.*, vol. 86, no. 2, pp. 1093-1109, Aug. 2021.

[34] T. Küstner et al., "CINENet: deep learning-based 3D cardiac CINE MRI reconstruction with multi-coil complex-valued 4D spatio-temporal convolutions," *Sci. Rep.*, vol. 10, no.1, p. 13710, Aug. 2020.

[35] P. Virtue, "Complex-valued deep learning with applications to magnetic resonance image synthesis," Ph.D. dissertation, University of California, Berkeley, Aug. 2018.

[36] V. Corona, A. Aviles-Rivero, N. Debroux, C. Le Guyader, and C. B. Schönlieb, "Variational multi-task MRI reconstruction: Joint reconstruction, registration and super-resolution," *Med. Image Anal.*, vol. 68, p. 101941, Feb. 2021.

[37] R. Shaul, I. David, O. Shitrit, and T. Riklin Raviv, "Subsampled brain MRI reconstruction by generative adversarial neural networks," *Med. Image Anal.*, vol. 65, p. 101747, Oct. 2020.

[38] J. Schlemper, J. Caballero, J. V. Hajnal, A. N. Price, and D. Rueckert, "A Deep Cascade of Convolutional Neural Networks for Dynamic MR Image Reconstruction," *IEEE Trans. Med. Imaging*, vol. 37, no. 2, pp. 491-503, Oct. 2018.

[39] S. Wang et al., "DIMENSION: Dynamic MR imaging with both k-space and spatial prior knowledge obtained via multi-supervised network training," *NMR Biomed.*, vol. 35, no. 4, p. e4131, Apr. 2022.

[40] O. Nitski, S. Nag, C. McIntosh, and B. Wang, "CDF-Net: Cross-Domain Fusion Network for Accelerated MRI Reconstruction," in *Medical Image Computing and Computer Assisted Intervention (MICCAI)*, 2020, pp. 421-430.

[41] C. M. Sandino, P. Lai, S. S. Vasanawala, J. Y. Cheng. "Accelerating cardiac cine MRI using a deep learning-based ESPIRiT reconstruction," *Magn. Reson. Med.*, vol. 85, no. 1, pp. 152-167, Jan. 2021.

[42] K. He, X. Zhang, S. Ren, and J. Sun, "Deep residual learning for image recognition," in *IEEE Conference on Computer Vision and Pattern Recognition (CVPR)*, 2016, pp. 770-778.

[43] G. Huang, Z. Liu, L. Van Der Maaten, and K. Q. Weinberger, "Densely connected convolutional networks," in *IEEE Conference on Computer Vision and Pattern Recognition (CVPR)*, 2017, pp. 2261-2269.

[44] Y. Chen, J. Li, H. Xiao, X. Jin, S. Yan, and J. Feng, "Dual path networks," in *Proceedings of the International Conference on Neural Information Processing Systems (NIPS)*, 2017, pp. 4470-4478.

[45] W. Wang, X. Li, T. Lu, and J. Yang, "Mixed link networks," in *International Joint Conference on Artificial Intelligence (IJCAI)*, 2018, pp. 2819-2825.

[46] W. Zhou, H. Du, W. Mei, and L. Fang, "Efficient structurally-strengthened generative adversarial network for MRI reconstruction," *Neurocomputing*, vol. 422, pp. 51-61, Jan. 2021.

[47] M. He, W. Zha, F. Tan, L. Rankine, S. Fain, and B. Driehuys, "A

comparison of two hyperpolarized $^{129}$Xe MRI ventilation quantification pipelines: The effect of signal to noise ratio," *Acad. Radiol.*, vol. 26, no. 7, pp. 949-959, Sep. 2019.

[48] M. Lustig and J. Pauly, "SPIRiT: Iterative Self-Consistent Parallel Imaging Reconstruction from Arbitrary k-Space Sampling," *Magn. Reson. Med.*, vol. 64, no. 2, pp. 457-71, Aug. 2010.

[49] X. Qu, W. Zhang, D. Guo, C. Cai, S. Cai, and Z. Chen, "Iterative thresholding compressed sensing MRI based on contourlet transform," *Inverse Probl. Sci. Eng.*, vol. 18, no. 6, pp. 737-758, Aug. 2010.

[50] Z. Wang, A. C. Bovik, H. R. Sheikh, and E. P. Simoncelli, "Image quality assessment: From error visibility to structural similarity," *IEEE Trans. Image Process.*, vol. 13, no. 4, pp. 600-612, Apr. 2004.

[51] G. Parraga, A. Ouriadov, A. Evans, and S. McKay, "Hyperpolarized $^3$He ventilation defects and apparent diffusion coefficients in chronic obstructive pulmonary disease," *Invest. Radiol.*, vol. 42, no. 6, pp. 384-391, Jun. 2007.

[52] L. Ebner et al., "Multireader Determination of Clinically Significant Obstruction Using Hyperpolarized $^{129}$Xe Ventilation MRI,". *Am. J. Roentgenology*, vol. 212, no. 4, pp. 758-165, Apr. 2019.

[53] P. P. Rebouças Filho, P. C. Cortez, A. C. da Silva Barros, V. H. Victor, and R. S. J. M. Tavares, "Novel and powerful 3D adaptive crisp active contour method applied in the segmentation of CT lung images," *Med. Image Anal.*, vol. 35, pp. 503-516, Jan. 2017.

[54] M. Kirby et al., "Hyperpolarized $^3$He Magnetic Resonance Functional Imaging Semiautomated Segmentation," *Acad. Radiol.*, vol. 19, no. 2, 141-152, Feb. 2012.

[55] J. Bertels, D. Robben, D. Vandermeulen, and P. Suetens, "Theoretical analysis and experimental validation of volume bias of soft Dice optimized segmentation maps in the context of inherent uncertainty," *Med. Image Anal.*, vol. 67, p. 101833, Jan. 2021.

[56] E. Shimron, J. I. Tamir, K. Wang, and M. Lustig, "Subtle inverse crimes: Naïvely using publicly available images could make reconstruction results seem misleadingly better!," in *International Society for Magnetic Resonance in Medicine (ISMRM)*, 2021.

[57] S. Wang, T. Zhou, and J. Bilmes, "Bias also matters: bias attribution for deep neural network explanation," in *International Conference on Machine Learning (ICML)*, 2019, pp. 6659-6667.